# High-Temperature Interfacial Antiferromagnetism in Graphene-Confined EuO/Co Heterostructures

Matteo Jugovac*,‡,⊥,±,†, Iulia Cojocariu‡,⊥,±,†, Gustav Bihlmayer‖, Tevfik Onur Menteş†, Andrea Locatelli†, Paolo Perna┼,*

⊥ Istituto di Struttura della Materia, Consiglio Nazionale delle Ricerche, Strada statale 14, km 163.5, 34149 Trieste, Italy

± Dipartimento di Fisica, Università degli Studi di Trieste, Via A. Valerio 2, 34127 Trieste, Italy

† Elettra Sincrotrone Trieste S.C.p.A., Strada statale 14, km 163.5, 34149 Trieste, Italy

‖ Peter Grünberg Institut (PGI-1), Forschungszentrum Jülich and JARA, 52425 Jülich, Germany

┼ IMDEA Nanociencia, C/Faraday 9, Campus de Cantoblanco, 28049 Madrid, Spain

* corresponding authors E-mail: matteojugovac@cnr.it; paolo.perna@imdea.org

‡ These authors contributed equally to this work.

ABSTRACT

Synthetic antiferromagnets combine antiparallel magnetic alignments with tunable functionality, yet realizing them in the ultimate two-dimensional limit remains a major challenge for spintronic, magnonic, and quantum technologies. Here, we demonstrate a route based on interfacial engineering: graphene confinement is used to grow an ordered, semiconducting EuO film of bilayer thickness at a buried EuO/Co interface, where antiferromagnetic coupling is stabilized by interfacial exchange. X-ray spectroscopy confirms that oxidation is selective for europium, while element-resolved magnetic imaging reveals an antiparallel alignment between EuO and Co domains that persists up to at least 483 K, which is far above the Curie temperature of bulk EuO, demonstrating a thermally robust, atomically thin antiferromagnetic building block. Angle-resolved photoemission shows that the graphene π bands are simultaneously strongly electron-doped and renormalized by the buried interface, in excellent agreement with dedicated first-principles calculations. Our results establish graphene-confined EuO/Co heterostructures as a versatile platform in which oxide chemistry, magnetic order, and graphene electronic states are engineered together, opening a route toward high-temperature two-dimensional synthetic antiferromagnets for ultrafast, ultralow-power spintronic and quantum communication technologies.

## INTRODUCTION

Graphene has become a versatile platform for engineering buried interfaces. Beyond its role as an atomically thin electronic material, it can act as a chemically robust overlayer that protects reactive species, mediates interactions with the underlying substrate, and confines atoms or compounds to a subsurface region only a few atomic layers thick.[1,2] Recent work on confinement heteroepitaxy has shown that the graphene/substrate interface can stabilize atomically thin crystalline phases that are not accessible as conventional free surfaces, including two-dimensional metals and ordered intercalated layers. In this configuration, graphene provides vertical confinement and environmental protection, while interactions with graphene and the supporting substrate can stabilize ordered ultrathin phases and modify their electronic structure through charge transfer, strain, hybridization, and interfacial registry.[3–6] The fact that such confined phases can remain structurally ordered makes graphene confinement a powerful route for designing buried two-dimensional heterostructures.

This concept has been developed mainly for non-magnetic intercalated metals and semimetals, where the confined layer is used to decouple graphene, modify its doping or strain, or generate new electronic phases.[3,4,7,8] However, graphene can also be an active layer in magnetic heterostructures. On ferromagnetic metals such as Co and Ni, graphene hybridizes with spin-polarized 3d states, leading to charge transfer, band renormalization, and proximity-induced magnetic effects.[9–12] At the same time, graphene has been used as an atomically thin spacer in synthetic magnetic heterostructures, where it can mediate interlayer exchange coupling between adjacent ferromagnetic layers.[13,14] These examples show that graphene is not only a passive cap but an active interfacial component whose electronic structure and coupling to neighboring materials can be used to control magnetic and electronic properties.[15–17]

Although pristine graphene is effectively impermeable, adspecies can enter the graphene/substrate interface through edges and defects under suitable thermal conditions. Once intercalated, they may form continuous metallic or compound layers confined between graphene and the substrate. Reactive intercalation can extend this approach to buried oxides, as demonstrated by the formation of interfacial $SiO_2$ beneath graphene.[18] Its extension to magnetic oxides, however, remains largely unexplored.

Europium monoxide is a particularly relevant oxide for magnetic applications. EuO is a prototypical ferromagnetic semiconductor, characterized by large $Eu^{2+}$ magnetic moments and an exchange-split conduction band below its Curie temperature.[19–21] This exchange splitting gives rise to spin-dependent electronic barriers, making EuO a model material for spin-selective transport and spin-filter tunnel barriers.[22] The ultrathin regime is particularly relevant for spin-filter applications because tunneling transmission decreases exponentially with barrier thickness. Precise thickness control at the nanometer scale is therefore essential to balance the magnitude and spin polarization of the tunneling current while preserving the chemical and magnetic integrity of EuO.[22–24] At the same time, this is precisely the regime in which EuO becomes most sensitive to stoichiometry, defects, oxidation state, and interface reactions.[23,24]

The interaction between EuO and graphene is also particularly attractive because ferromagnetic insulators can induce spin-dependent effects in graphene without short-circuiting its transport channel. First-principles studies of graphene interfaced with bulk EuO have predicted exchange splitting of the graphene bands,[25,26] while experiments on EuO/graphene have reported signatures of proximity-induced magnetism, anomalous Hall response, and strong sensitivity to synthesis conditions and interfacial doping.[27,28] These results underscore the importance of precise control of the EuO layer structure and thickness, as well as its interfacial interactions. In most previous EuO/graphene studies, EuO was grown on top of graphene as a relatively thick film, whereas the controlled formation of a stoichiometric EuO layer beneath graphene, in a truly confined geometry, remains largely unexplored.

Interfacing EuO with cobalt provides a means of exploiting interfacial exchange in the ultrathin limit. Calculations for isolated EuO(100) films predict a strong thickness dependence of the intrinsic Curie temperature, decreasing from the bulk value to approximately 15 K in the monolayer limit.[29]. Direct contact with a ferromagnetic metal, however, provides an interfacial exchange field that can control and stabilize the magnetic response of EuO.[30] The bilayer limit is therefore not only a structural feature but a central part of the physics: most of the EuO layer lies within the interfacial exchange region, allowing the cobalt substrate to control the EuO magnetic configuration.[30] Combining EuO with ferromagnetic layers beneath graphene thus provides a way to test whether graphene confinement can stabilize an ultrathin magnetic oxide while preserving both exchange coupling to a ferromagnetic metal and electronic functionality in the graphene layer.

Here, we investigate a graphene/EuO/Co heterostructure in which europium is intercalated beneath graphene and selectively oxidized to form a nominally two-layer-thick EuO layer at the cobalt interface. This geometry combines graphene confinement of a reactive ultrathin oxide with exchange coupling to cobalt, which stabilizes the EuO magnetic alignment. By combining diffraction, spectroscopy, microscopy, angle-resolved photoemission, and first-principles calculations, we examine the structural, chemical, electronic, and magnetic properties of the system. We show that graphene confinement stabilizes an ordered EuO bilayer at the Co interface, that EuO and Co domains align antiparallel up to 483 K, and that the graphene $\pi$ bands remain visible but are strongly modified by the buried oxide/metal interface, as also confirmed from density functional theory calculations. In contrast to previous EuO/graphene architectures, the present heterostructure features an ordered EuO bilayer beneath graphene that is directly exchange-coupled to a ferromagnetic metal. The key advance is the simultaneous demonstration of graphene-confined oxide formation, antiparallel EuO/Co alignment persisting to 483 K, and substantial modification of the graphene π bands within the same buried interface.

## RESULTS AND DISCUSSION

The fabrication of graphene/EuO/cobalt heterostructures is carried out under ultrahigh-vacuum conditions, while *in situ* monitoring the surface structure by low-energy electron diffraction (LEED). Further details on the growth protocol are presented in the Methods Section. The 35 eV diffraction images in Figures 1a-d capture the most relevant stages of the preparation protocol, beginning with the pristine graphene/Co interface. At this stage, the pattern is dominated by a characteristic moiré modulation around the specular reflection (Figure 1a), arising from the lattice mismatch between the ultrathin Co film and the Re(0001) substrate.[31] The relative alignment between graphene and Co is confirmed by the LEED pattern in Supporting Information Figure S1a, which shows an epitaxial $(1 \times 1)$ registry between the two lattices.

Upon deposition of one-third of a monolayer of Eu onto graphene/Co at approximately 400 K, Eu orders on top of graphene in a $\sqrt{3} \times \sqrt{3}R30°$ superstructure, consistent with an $EuC_6$-type phase. Annealing to 540 K then drives Eu intercalation beneath graphene, while retaining the same periodicity at the graphene/Co interface (Figure 1b).[32] To induce oxide formation, the Eu-intercalated graphene/Co interface is exposed to molecular oxygen at about 470 K. This leads to the disappearance of the interfacial reconstruction, indicating that the Eu phase loses its long-range order as europium reacts with oxygen. Long-range order is restored upon deposition of an additional one-third of a monolayer of Eu at 540 K, giving rise to a complex LEED pattern comprising sharp first-order spots from three rotational domains of EuO(100) (Figure 1d). Analysis of their reciprocal-space vectors yields a EuO lattice parameter of 5.125 Å, only about 0.3% smaller than the bulk value (5.14 Å).

The evolution of the interfacial ordering is quantified in Figure 1e through azimuthal intensity profiles extracted at the reciprocal-space radii of the corresponding diffraction maxima in Figures 1b–d. The intercalated Eu phase gives sharp maxima separated by 60° (blue curve), as expected for the $\sqrt{3} \times \sqrt{3}R30°$ reconstruction. After oxidation, these maxima broaden and additional peaks appear at 30° offsets (red curve), revealing a different oxide-related structure with reduced azimuthal coherence. This broad angular distribution indicates that europium oxide initially nucleates as small crystallites with partial rotational disorder. Following the second Eu deposition, the intensity is concentrated again into sharp and intense maxima, separated by 30° (green curve), confirming the formation of ordered EuO rotational domains, consistent with the LEED pattern in Figure 1d.

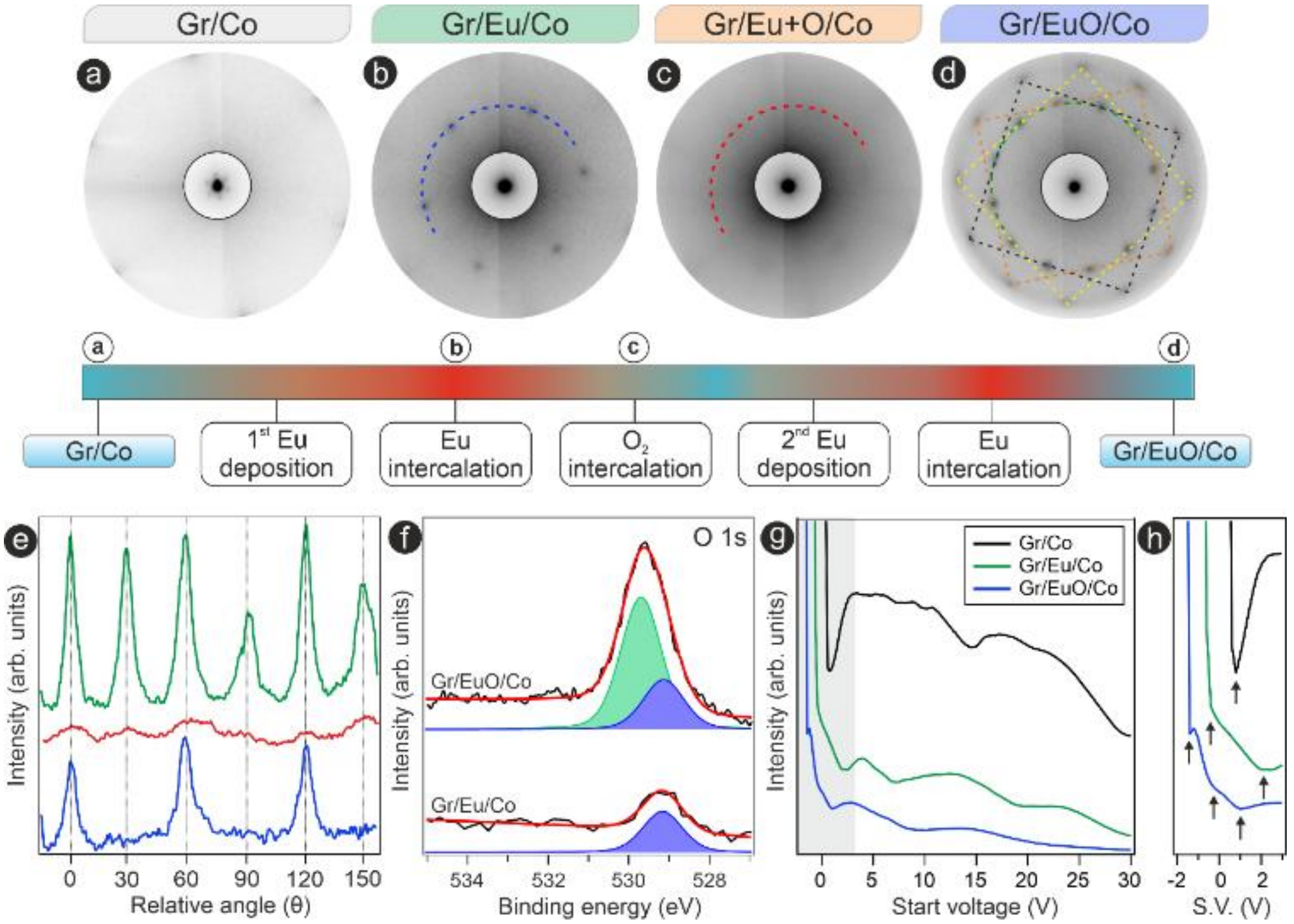

**Figure 1. Structural and chemical evolution during the formation of graphene-confined EuO.** a–d) *LEED patterns acquired at 35 eV during the preparation sequence: a) pristine graphene/Co; b) Eu-intercalated graphene/Co; c) after oxygen exposure; and d) final graphene/EuO/Co heterostructure. Dashed black, yellow and orange lines indicate the unit cells of EuO(100) domains. e) Azimuthal intensity profiles extracted at the reciprocal-space radii of the corresponding diffraction maxima, along the dashed paths in panels b–d, showing the transition from the ordered Eu-intercalated phase to the oxidized intermediate structure and finally to ordered EuO domains. The profiles are normalized to their respective maxima and vertically offset for clarity. f) O* 1*s XPS spectra (hν = 650 eV) acquired before and after EuO formation. g) LEEM-IV reflectivity spectra recorded during the preparation sequence. Arrows indicate the positions of dips in each spectrum. h) Enlarged view of the low-start-voltage region, with arrows marking the reflectivity minima.*

The chemical evolution of the confined Eu layer is monitored by O 1s X-ray photoelectron spectroscopy (XPS) (Figure 1f). Before oxidation, only weak O 1s intensity is detected near 529.2 eV, indicating a small amount of residual oxygen-containing species.[8] After oxygen exposure and EuO formation, the O 1s peak shifts to 529.7 eV, in agreement with reported values for lattice oxygen in EuO films.[20] The absence of pronounced high-binding-energy components indicates that hydroxyl species or strongly over-oxidized oxygen environments are negligible under the present growth conditions. Together with the $Eu^{2+}$ fingerprint obtained from Eu M-edge X-ray absorption spectroscopy (XAS), discussed below, the O 1s spectra support the formation of predominantly divalent EuO phase.

LEEM-IV spectroscopy provides a complementary fingerprint of the electronic changes accompanying Eu intercalation and oxidation (Figure 1g,h). Relative to graphene/Co, Eu intercalation introduces a second reflectivity minimum and shifts the mirror-mode transition by 1.17 V toward lower bias. This shift is consistent with electron donation from intercalated Eu to graphene and a substantial reduction in the local work function.[32] In the final graphene/EuO/Co structure, the onset downshifts by a further 0.46 V and the reflectivity develops additional weak features, confirming that EuO formation modifies the unoccupied interfacial electronic structure. Taking pristine graphene/Co as the reference, the total shift of 1.63 V corresponds to an estimated work function of approximately 2.20 eV for the final heterostructure, consistent with the low-work-function character of EuO-containing surfaces.[33] The calibrated Eu dose corresponds nominally to two EuO atomic layers. Together with the absence of lateral contrast associated with different thicknesses in LEEM, this suggests that the confined oxide has a laterally homogeneous, nominal two-layer thickness.[34]

X-ray absorption at the Co and Eu edges confirms that oxide formation is confined to europium while the cobalt layer remains metallic. The Co $L_{32}$-edge spectrum retains the characteristic line shape of metallic cobalt (Figure 2a). In particular, the absence of the multiplet features typical of Co oxides indicates that the Co film remains predominantly metallic after Eu intercalation and oxidation.[35] At the Eu $M_{5,4}$ edges, the absorption profile shows a well-defined multiplet structure characteristic of divalent europium, with the $M_5$ maximum located at about 1128 eV (Figure 2b), in agreement with $Eu^{2+}$ in EuO thin films.[36] No additional resonances are detected at higher photon energies, where $Eu^{3+}$ oxides such as $Eu_2O_3$ are found (see over-oxidation experiment in the Supporting Information Figure S2). This confirms the divalent character of the confined EuO layer and is consistent with the O 1s XPS spectra discussed above. Graphene confinement thus enables selective oxidation of the intercalated Eu layer while protecting Co from significant oxidation.

To assess the magnetic changes induced by EuO formation, we first characterize the magnetic domain structure of pristine Gr/Co/Re(0001) before Eu deposition. At room temperature and zero applied magnetic field, XMCD-PEEM imaging at the Co $L_3$ absorption maximum ($h\nu = 777.25$ eV) reveals a micrometer-scale, labyrinthine-type domain pattern in the graphene-covered 5-ML Co film (Figure S1b,c), typical of ultrathin cobalt covered by a graphene layer.[37,38] Against this reference, the magnetic configuration of the buried Gr/EuO/Co interface is examined by XMCD-PEEM. Helicity-dependent images were acquired at the Co and Eu absorption maxima, $h\nu = 777.25$ eV and $1128.50$ eV, respectively, in the as-grown state and at zero applied magnetic field. The photon beam impinges at an angle of 16° relative to the sample surface, making the XMCD contrast predominantly sensitive to the in-plane magnetization component parallel to the beam projection. The magnetic contrast images in Figures 2c-h are obtained from the normalized difference between images recorded with opposite photon helicities. Across the investigated temperature range, from 156 to 483 K, the Co and Eu images exhibit inverted contrast within their respective magnetic domains. The antiparallel alignment of the projected EuO and Co magnetization components demonstrates antiferromagnetic interfacial coupling between EuO and Co (see Figure S3 for the spatial comparison). The persistence of this correlation up to 483 K indicates an exchange-stabilized interfacial magnetic configuration rather than bulk-like EuO ferromagnetism with $T_C = 69$ K.[20] We note that XMCD-PEEM measurements of the overoxidized $Eu_2O_3$ interfacial phase reveal no detectable magnetic contrast within the experimental sensitivity, consistent with the $J = 0$ ground state of $Eu^{3+}$ (Figure S2b,c).

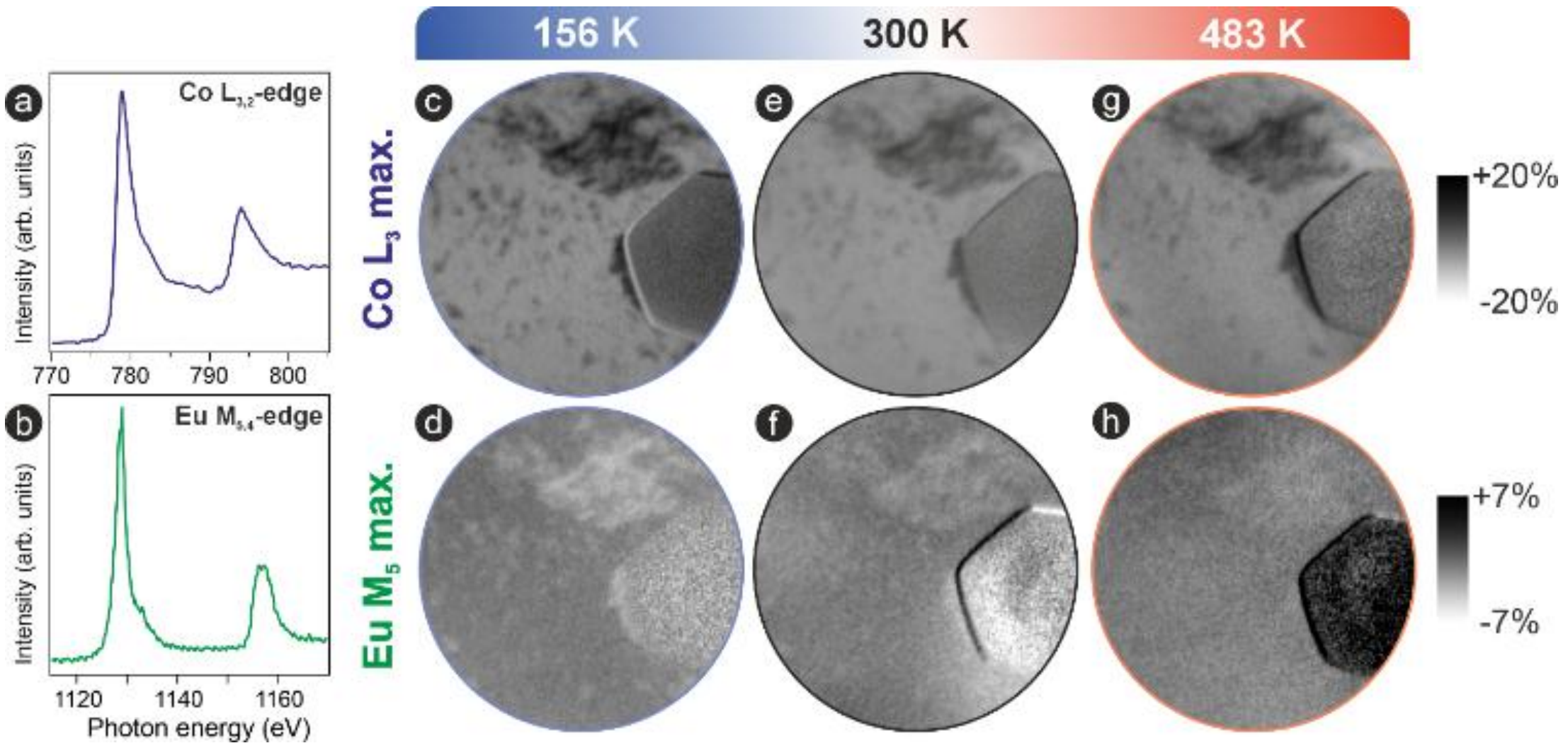

***Figure 2. Element-resolved spectroscopy and magnetic imaging of graphene/EuO/Co.*** *a) Co $L_{3,2}$-edge X-ray absorption spectrum of the graphene/EuO/Co heterostructure, displaying the characteristic line shape of metallic cobalt. b) Eu $M_{5,4}$-edge*

*X-ray absorption spectrum, showing the multiplet structure characteristic of divalent Eu in EuO. c–h) XMCD-PEEM images acquired at the Co $L_3$ and Eu $M_5$ edges at 156 K, 300 K, and 483 K. Images were recorded in the as-grown state, at zero field and calculated as the normalized difference $(I^+-I^-)/(I^++I^-)$. The inverted magnetic contrast between the Co and Eu images in the same sample regions indicates antiparallel alignment of the EuO and Co magnetic domains. The large faceted feature on the right is a substrate defect used as a reference for spatial orientation and drift correction of the Co and Eu images.*

Antiparallel magnetic alignment driven by interfacial exchange has been reported in EuO/ferromagnet systems and is closely related to the broader concept of synthetic antiferromagnetic coupling.[30,39] In the present heterostructure, the persistence of antiparallel EuO/Co contrast up to at least 483 K points to a strong exchange pathway across the buried interface. This behavior is consistent with recent magnetic depth-profiling studies of Fe/EuO and Co/EuO bilayers, which showed that antiferromagnetic proximity coupling can stabilize EuO magnetic order above its bulk $T_C$, but only over a short interfacial length scale.[30] The effect is therefore strongest in the ultrathin limit, where most of the EuO layer lies within the proximity-coupled region.

The confined EuO layer also reshapes the electronic structure of the graphene/Co interface, as revealed by ARPES. Figure 3a shows the photoemission intensity map acquired along the $\overline{\mathrm{M}\Gamma\mathrm{K}}$ direction of the first Brillouin zone. The spectrum is dominated by weakly dispersive features centered at binding energies of approximately 2.0 eV and 4.9 eV, assigned to Eu 4f-derived and O 2p-derived states, respectively. Their limited dispersion and energy position are consistent with the localized character of Eu 4f electrons and the broader O 2p manifold reported for EuO thin films and related systems.[21] A weaker parabolic contribution associated with the graphene π bands is also visible. Its reduced intensity relative to the Eu- and O-derived states is attributed to the lower photoemission cross-section of C 2p orbitals at the selected photon energy.[40] No distinct Co-derived dispersion is resolved in the ARPES map, consistent with attenuation of the buried Co contribution by the EuO and graphene layers.

To enhance the graphene-derived contribution, each constant-energy intensity profile is normalized to its maximum value. (Figure 3b). This reveals well-defined $\pi$ and $\pi^*$ bands separated by a region of strongly reduced spectral weight. The $\pi$ band minimum at $\overline{\Gamma}$ is located at a binding energy of 9.46 eV, consistent with strongly interacting graphene systems and with ARPES measurements of graphene on transition-metal substrates.[7,11] The separation of the $\pi$ and $\pi^*$ bands can arise from broken sublattice equivalence in graphene and from hybridization with the confined interfacial layer. In the present system, its energetic overlap with the Eu manifold suggests that interaction between graphene and Eu-derived states contributes to the modification of the Dirac cone.[32] Nevertheless, ARPES alone does not uniquely distinguish an intrinsic band gap from an avoided crossing or a matrix-element-induced suppression of one of the spectral branches.

A closer view of the near-$E_F$ region is shown in Figure 3c, where the ARPES cut is centered at the $\overline{\mathrm{K}}$ point. On either side of the region of reduced spectral weight, the $\pi$ and $\pi^*$ branches retain an approximately linear dispersion. Linear fits yield Fermi velocities of $v_h = (8.61 \pm 0.32) \times 10^5$ m/s on the hole side and $v_e = (7.89 \pm 0.32) \times 10^5$ m/s on the electron side, corresponding to an average value of $v_F^{avg} = 8.25 \times 10^5$m/s. These values are slightly smaller than those of freestanding graphene but remain within the range reported for graphene on metallic substrates and intercalated interfaces.[2,41] The approximately 9% electron–hole asymmetry further indicates a departure from ideal Dirac behavior, consistent with many-body renormalization and substrate-induced effects observed in supported graphene.[41]

The extracted Fermi velocities correspond to an effective nearest-neighbor hopping parameter of $t_{eff} \approx 2.55$ eV. These values are slightly smaller than the 2.7 eV value generally reported for freestanding graphene. Such a reduction is consistent with substrate-induced renormalization of the $\pi$-electron system.[2] The linear extrapolations give a separation of 0.67 eV between the $\pi$ and $\pi^*$ branch onsets. We therefore refer to this value as an apparent spectral gap, because the present measurement cannot uniquely distinguish a true quasiparticle gap from an avoided crossing or suppressed spectral weight. The midpoint of the two extrapolated onsets lies at a binding energy of approximately 1.60 eV and is taken as the effective Dirac-point energy. This large shift indicates strong n-type doping. Within the linear-cone approximation, the corresponding carrier density is $n = 2.7 \times 10^{14}$ cm$^{-2}$, equivalent to approximately 0.07 electrons per carbon atom, comparable to heavily doped graphene.[42,43]

The Fermi surface map in Figure 3d corroborates the picture outlined above. Six well-defined electron pockets are observed around the $\overline{\mathrm{K}}$ points, consistent with the downward shift of the Dirac point and the strong electron doping.

Additional intensity is detected near $\overline{\mathrm{M}}$, where the graphene $\pi^*$ dispersion contains a relatively flat saddle-point region.[32,42] This observation is consistent with substantial electron doping that brings the $\pi^*$ states at $\overline{\mathrm{M}}$ closer to $E_F$, although the measured intensity does not establish that the van Hove singularity itself is occupied.[7]

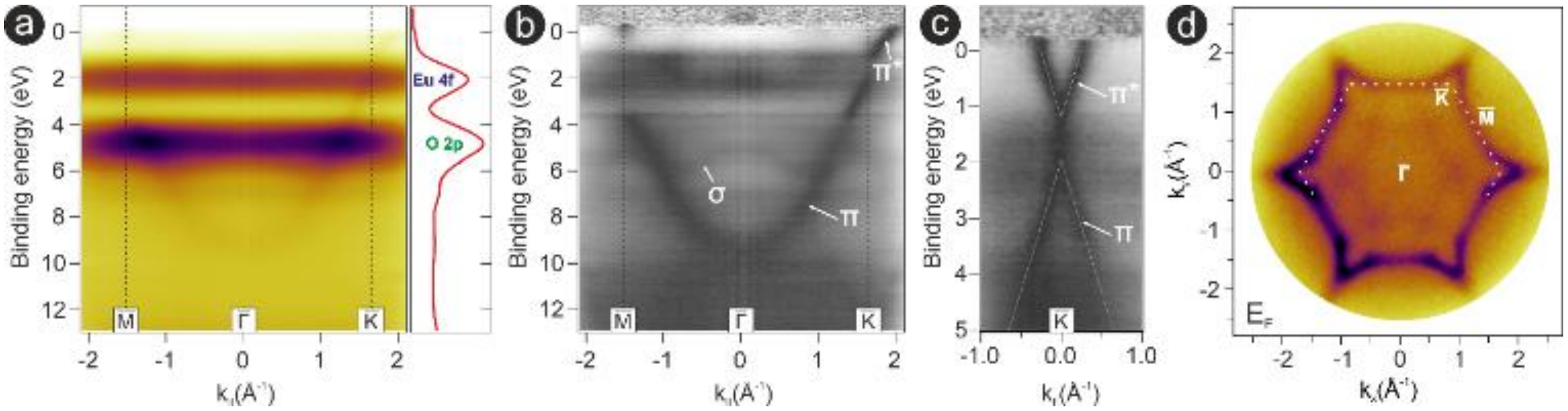

***Figure 3. ARPES characterization of the graphene/EuO/Co electronic structure.*** *a) ARPES intensity map acquired along the $\overline{M\Gamma K}$ direction, along with the integrated energy distribution curve on the right side. b) Normalized ARPES intensity map highlighting the graphene $\pi$ and $\pi^*$ bands. c) High-resolution ARPES cut around the $\overline{K}$ point. Linear fits to the $\pi$ and $\pi^*$ branches are marked by dashed white lines. d) 2D momentum map acquired at the Fermi level, along with the sketch of the surface Brillouin zone as a white dashed line.*

The electronic structure is calculated for the lowest-energy graphene/EuO/Co(0001) model, in which the EuO magnetization is antiparallel to that of Co. This magnetic configuration is energetically favored over the parallel state. The structural model contains two confined EuO layers between graphene and cobalt and corresponds to the lowest-energy registry among the configurations considered. Figure 4 shows the spin-resolved projected band structure, in which the majority- and minority-spin contributions of graphene (blue/red), europium (orange/green), and oxygen (black/brown) are represented separately. To account for the strongly localized Eu 4f electrons, the calculations are performed within the DFT+U framework using an effective Hubbard parameter $\mathrm{U} = 6$ eV.[44,45]

The calculated band structure captures the main features observed experimentally. In particular, the Eu-derived states appear as weakly dispersive bands in the same binding-energy region as the experimental Eu feature. The oxygen-derived states form a broader manifold at higher binding energy, consistent with the characteristic distribution of O 2p states in EuO.[21] Although the absolute energies are slightly shifted with respect to the ARPES spectra, the relative ordering and orbital character of the Eu 4f and O 2p states are reproduced.

The graphene-projected $\pi$ bands remain identifiable near the $\overline{\mathrm{K}}$ point as approximately linear branches, although hybridization produces deviations from an ideal Dirac cone and redistributes their spectral weight, consistently with the experimental ARPES results. At the $\overline{\mathrm{M}}$ point, the $\pi^*$ band lies slightly above the Fermi level, consistent with strong electron doping and the experimentally observed spectral weight in this region. Only a weak spin polarization of the graphene $\pi$ states is obtained, without a pronounced exchange splitting. The smaller splitting compared with earlier predictions for idealized graphene/EuO interfaces[25,26] likely arises from the combined effects of atomic registry and Co-EuO hybridization, which modifies the electronic structure of the confined EuO bilayer and its coupling to graphene.

In the calculation, the Eu atoms carry magnetic moments close to 7 $\mu_B$, consistent with the $Eu^{2+}$ $4f^7$ configuration. The Co atoms retain local moments comparable to those of metallic ultrathin Co films, indicating that the EuO layer does not strongly influence the Co magnetization at the interface. We explored different structural models for the EuO bilayer; the energetically preferred configuration consists of two buckled honeycomb layers (buckling of about 0.6 Å, with O oriented towards the substrate) separated by 2.38 Å. In contrast to bulk EuO, the distance between the Eu and O (111) planes is reduced to lower the dipole moment, and Eu and O atoms are stacked vertically (see Supporting Information Figure S4). The EuO layer retains ferromagnetic alignment among Eu moments, while its net magnetization couples antiferromagnetically to Co. The antiparallel EuO/Co configuration is lower in energy by 27 meV per Eu atom compared with the parallel alignment.

The effect of overoxidation is further examined by ARPES on the $Eu_2O_3$-rich interfacial phase (Supporting Information

Figure S5). The wide-range spectrum shows a broad valence-band manifold at binding energies of approximately 4-7 eV, while the pronounced $Eu^{2+}$ $4f$ feature observed near 2 eV for the EuO interface is absent. Around $\bar{K}$, the graphene $\pi$ and $\pi^*$ branches remain visible and are separated by approximately 0.46 eV. The midpoint of the two branch onsets lies at a binding energy of approximately 1.39 eV and is taken as the effective Dirac-point energy. Graphene, therefore, remains strongly n-doped, although the smaller Dirac-point shift relative to the EuO interface indicates reduced electron transfer upon overoxidation. Consistently, the Fermi-level momentum map displays separated electron pockets around the $\bar{K}$ points, but no appreciable intensity near $\bar{M}$, indicating that the $\pi^*$ states do not approach the $\bar{M}$-point saddle region as closely as in graphene/EuO/Co. Together with the $Eu^{3+}$ XAS fingerprint and the absence of detectable XMCD contrast, these results demonstrate that overoxidation modifies both the charge transfer to graphene and the magnetic response of the confined interface.

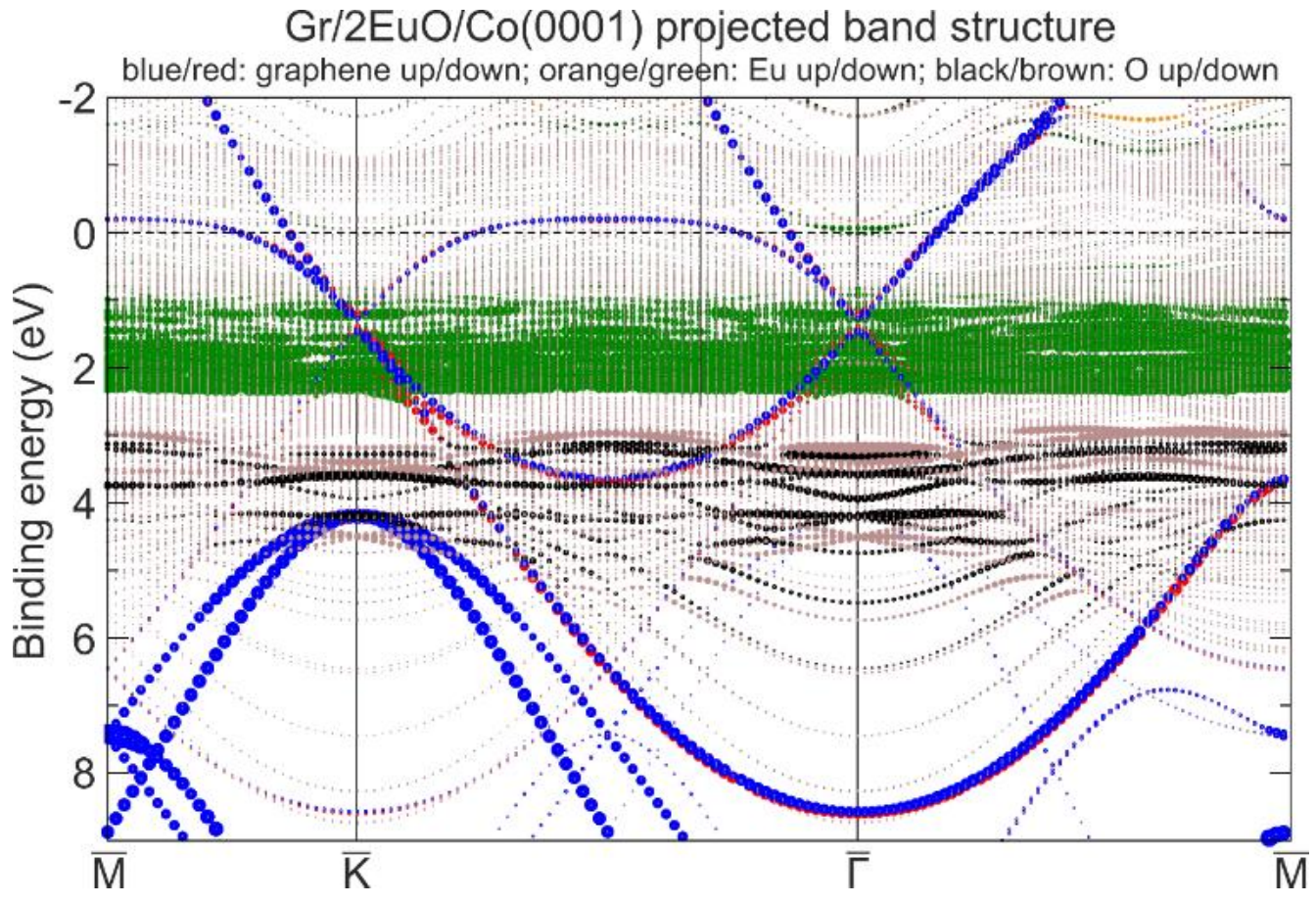

***Figure 4. Calculated electronic and magnetic structure of graphene/2(EuO)/Co.*** *Spin- and atom-projected band structure calculated for antiparallel alignment between bilayer EuO and cobalt. The majority- and minority-spin contributions of graphene are shown in blue and red, those of Eu in orange and green, and those of O in black and brown.*

Taken together, these results identify graphene/EuO/Co as an exchange-coupled oxide/metal interface confined by a two-dimensional material. The persistence of antiparallel EuO and Co magnetic domains up to at least 483 K is consistent with proximity-coupled Fe/EuO and Co/EuO bilayers, where EuO magnetic order is stabilized above the bulk $T_C$ over a short interfacial length scale.[30] The nominal two-layer thickness is therefore central to the present system, because most of the EuO layer lies within the exchange-coupled region. The antiparallel alignment also situates the system within the broader context of synthetic antiferromagnetic heterostructures, in which antiparallel magnetic configurations are engineered via interfacial exchange rather than intrinsic antiferromagnetism.[13,14] Here, the antiparallel alignment is imposed at the EuO/Co interface, while graphene confines the EuO formation and preserves detectable Dirac-like π bands that become strongly electron-doped and renormalized by the buried interface. In this geometry, the magnetic and electronic responses are both governed by the buried interface: Co controls the EuO alignment, while the confined oxide strongly reshapes the graphene π bands.

## CONCLUSION

We have shown that graphene confinement enables the formation of an ordered, predominantly divalent EuO film with a nominal thickness of two atomic layers at the graphene/cobalt interface, as identified by diffraction, spectroscopic, and photoemission fingerprints. Beyond stabilizing the confined oxide, graphene acts as a selective reaction membrane: it allows oxidation of the intercalated Eu while preserving the metallic character of the underlying cobalt. This two-layer limit is central to the system. Because the oxide is entirely interfacial, its magnetic response is governed by exchange coupling to cobalt rather than by bulk-like EuO ferromagnetism. This coupling imposes an antiparallel EuO/Co magnetic configuration that persists up to at least 483 K (far above the Curie temperature of bulk EuO) establishing graphene-confined EuO/Co as a thermally robust, atomically thin synthetic antiferromagnet. First-principles calculations support this interpretation by identifying the antiparallel state as energetically favored. At the same time, EuO formation strongly modifies graphene: the π bands remain visible but become heavily electron-doped and display strongly modified spectral weight around the Dirac point, demonstrating that the buried oxide can be used to engineer the graphene electronic structure in situ.

Graphene/EuO/Co therefore provides a versatile confined-oxide heterostructure in which oxide chemistry, interfacial magnetic order, and graphene electronic states are engineered together. These results establish graphene-confined EuO/Co as a model platform for investigating exchange-stabilized oxide magnetism at the atomic limit, and suggest a general route, i.e., graphene-confined reactive oxides coupled to ferromagnets, toward high-temperature two-dimensional synthetic antiferromagnets for spin-selective transport, magnonics, and quantum technologies. Future work exploiting the spin-selective character of the EuO conduction band and the gate-tunable graphene doping could enable direct control of interfacial spin alignment in operando device geometries.

## METHODS

*Sample preparation:* All experiments were performed under ultrahigh-vacuum conditions, with a base pressure below $2 \times 10^{-10}$ mbar in the preparation and analysis chambers. The substrate used for the growth was a Re(0001) single crystal. Prior to film growth, the substrate was cleaned by repeated annealing cycles in molecular oxygen, followed by high-temperature flashes in ultrahigh vacuum, until a sharp $(1 \times 1)$ LEED pattern was obtained.

Graphene was prepared in two steps. First, nanometer-sized graphene islands were synthesized directly on Re(0001) by ethylene chemical vapor deposition. Cobalt was then deposited onto the partially graphene-covered Re(0001) surface from a 99.995% purity Co rod using an electron-beam evaporator. The deposition was carried out at 520 K up to a thickness of 5 monolayers. After Co deposition, the graphene growth was completed by further ethylene chemical vapor deposition at 670 K, yielding the graphene/Co interface. The structural quality and cleanliness of the sample were checked by LEED and X-ray photoelectron spectroscopy.

Europium was evaporated from a homemade evaporator at a rate of approximately 0.65 Å $min^{-1}$. One-third of a monolayer of Eu was deposited onto graphene/Co at approximately 400 K, leading to the formation of a $\sqrt{3} \times \sqrt{3}R30°$ reconstruction. The sample was then annealed to 540 K to promote the intercalation of Eu at the graphene/Co interface.

For EuO formation, the Eu-intercalated graphene/Co sample was exposed to molecular oxygen at $1.5 \times 10^{-9}$ mbar and at approximately 470 K to promote oxygen intercalation beneath graphene.[8] After oxidation, an additional one-third of a monolayer of Eu was deposited at 540 K, yielding the graphene-confined EuO layer. The sample temperature was measured using a K-type thermocouple.

*Experimental methods:* LEEM, LEED, XMCD-PEEM, and ARPES measurements were carried out at the Nanospectroscopy beamline of the Elettra Synchrotron in Trieste, Italy, using the Spectroscopic Photoemission and Low Energy Electron Microscope (Elmitec GmbH, SPELEEM III) operated in ultrahigh vacuum.[46–48] The microscope can operate in real-space LEEM mode and reciprocal-space LEED mode. In this instrument, the electron energy is selected by applying a voltage bias to the sample, commonly referred to as the start voltage. Energy filtering is provided by a hemispherical electron energy analyzer (Elmitec R200), providing an energy resolution as good as 60 meV in microspot-spectroscopy mode and 100 meV in both imaging and diffraction mode. The photon beam impinged at 16° relative to the sample surface in both XAS and ARPES measurements.

LEEM was used to image the sample surface and to acquire intensity-voltage reflectivity curves. LEED was used to follow the evolution of the surface and interfacial diffraction patterns during the preparation sequence. When coupled to the synchrotron X-ray source with tunable photon energy and polarization, the SPELEEM enables chemical and magnetic imaging in real space via XPEEM and XMCD-PEEM, as well as band-structure mapping in reciprocal space through ARPES.

ARPES measurements were performed in reciprocal-space mode using photons with an energy of 40 eV and vertical polarization (p-polarization). Magnetic contrast images were acquired using XMCD-PEEM at the Co $L_3$ and Eu $M_5$ absorption-edge maxima by taking the difference between two images recorded with opposite photon helicities and normalizing to their sum.

*First-principles calculations:* First-principles calculations were performed within density functional theory using the generalized gradient approximation and including van der Waals corrections in the D3 form.[49,50] The calculations were carried out using the full-potential linearized augmented-plane-wave method in thin-film geometry,[51] as implemented in the FLEUR code.[52] To describe the localized Eu 4f electrons, calculations were performed within the DFT+U framework using an effective Hubbard parameter of $U = 6$ eV. Spin–orbit coupling was included self-consistently in the band-structure calculations.

The graphene/EuO/Co heterostructure was modeled using a thin-film geometry with a cobalt slab, two confined EuO layers, and a graphene overlayer. Several relative registries of the confined layers were examined, and the lowest-energy configuration was adopted for the electronic-structure analysis. The magnetic configuration was evaluated by comparing ferromagnetic and antiferromagnetic alignment of EuO relative to cobalt. The projected band structures were resolved into orbital and spin contributions from graphene, europium, oxygen, and cobalt, allowing direct comparison with the experimental ARPES spectra.

ACKNOWLEDGEMENTS

G.B. gratefully acknowledges computing time on the supercomputer JURECA-DC at Forschungszentrum Jülich. P.P. acknowledges support by the Spanish AEI through Projects PID2024-157112OB-C53 (HYBRIDOS: HYPERFAN), M-ERANET 2024 PCI2025-163177 (GREEN-MEM) and the "Severo Ochoa" Programme for Centres of Excellence in R&D (CEX2025-001660-S), and by the Community of Madrid through the programs Mag4TIC-CM (TEC280-CM).

BIBLIOGRAPHY

[1] K. S. Novoselov, A. K. Geim, S. V. Morozov, D. Jiang, Y. Zhang, S. V. Dubonos, I. V. Grigorieva, A. A. Firsov, *Science* **2004**, *306*, 666.

[2] A. H. Castro Neto, F. Guinea, N. M. R. Peres, K. S. Novoselov, A. K. Geim, *Rev. Mod. Phys.* **2009**, *81*, 109.

[3] N. Briggs, B. Bersch, Y. Wang, J. Jiang, R. J. Koch, N. Nayir, K. Wang, M. Kolmer, W. Ko, A. De La Fuente Duran, S. Subramanian, C. Dong, J. Shallenberger, M. Fu, Q. Zou, Y.-W. Chuang, Z. Gai, A.-P. Li, A. Bostwick, C. Jozwiak, C.-Z. Chang, E. Rotenberg, J. Zhu, A. C. T. Van Duin, V. Crespi, J. A. Robinson, *Nat. Mater.* **2020**, *19*, 637.

[4] Z. Mamiyev, N. Tilgner, N. O. Balayeva, D. R. T. Zahn, T. Seyller, C. Tegenkamp, *npj 2D Mater Appl* **2026**, *10*, 51.

[5] V. D. Pham, C. Dong, J. A. Robinson, *Nanoscale Adv.* **2023**, *5*, 5601.

[6] J. Halle, N. Néel, J. Kröger, *Langmuir* **2019**, *35*, 2554.

[7] M. Jugovac, C. Tresca, I. Cojocariu, G. Di Santo, W. Zhao, L. Petaccia, P. Moras, G. Profeta, F. Bisti, *Physical Review B* **2022**, *105*, L241107.

[8] M. Jugovac, F. Genuzio, T. O. Menteş, A. Locatelli, G. Zamborlini, V. Feyer, C. M. Schneider, *Carbon* **2020**, *163*, 341.

[9] G. Giovannetti, P. A. Khomyakov, G. Brocks, V. M. Karpan, J. Van Den Brink, P. J. Kelly, *Phys. Rev. Lett.* **2008**, *101*, 026803.

[10] P. A. Khomyakov, G. Giovannetti, P. C. Rusu, G. Brocks, J. Van Den Brink, P. J. Kelly, *Phys. Rev. B* **2009**, *79*, 195425.

[11] A. Varykhalov, O. Rader, *Phys. Rev. B* **2009**, *80*, 035437.

[12] H. Haugen, D. Huertas-Hernando, A. Brataas, *Phys. Rev. B* **2008**, *77*, 115406.

[13] P. Gargiani, R. Cuadrado, H. B. Vasili, M. Pruneda, M. Valvidares, *Nat Commun* **2017**, *8*, 699.

[14] C. A. Brondin, M. Hsouna, F. Genuzio, M. Jugovac, M. Zając, E. Partyka-Jankowska, S. Bonetti, A. Locatelli, N. Stojić, T. O. Menteş, *Carbon* **2026**, *249*, 121169.

[15] F. Ajejas, A. Gudín, R. Guerrero, A. Anadón Barcelona, J. M. Diez, L. de Melo Costa, P. Olleros, M. A. Niño, S. Pizzini, J. Vogel, M. Valvidares, P. Gargiani, M. Cabero, M. Varela, J. Camarero, R. Miranda, P. Perna, *Nano Lett.* **2018**, *18*, 5364.

[16] M. Blanco-Rey, P. Perna, A. Gudin, J. M. Diez, A. Anadón, P. Olleros-Rodríguez, L. de Melo Costa, M. Valvidares, P. Gargiani, A. Guedeja-Marron, M. Cabero, M. Varela, C. García-Fernández, M. M. Otrokov, J. Camarero, R. Miranda, A. Arnau, J. I. Cerdá, *ACS Appl. Nano Mater.* **2021**, *4*, 4398.

[17] A. Anadón, A. Pezo, I. Arnay, R. Guerrero, A. Gudín, A. Guio, M. Yactayo, J. Ghanbaja, J. Camarero, A. Manchon, S. Petit-Watelot, P. Perna, J. Rojas-Sánchez, *Advanced Materials* **2025**, 2418541.

[18] S. Lizzit, R. Larciprete, P. Lacovig, M. Dalmiglio, F. Orlando, A. Baraldi, L. Gammelgaard, L. Barreto, M. Bianchi, E. Perkins, P. Hofmann, *Nano Lett.* **2012**, *12*, 4503.

[19] A. Mauger, C. Godart, *Physics Reports* **1986**, *141*, 51.

[20] P. G. Steeneken, L. H. Tjeng, I. Elfimov, G. A. Sawatzky, G. Ghiringhelli, N. B. Brookes, D.-J. Huang, *Phys. Rev. Lett.* **2002**, *88*, 047201.

[21] T. Heider, T. Gerber, O. Köksal, M. Eschbach, E. Młyńczak, P. Lömker, P. Gospodaric, M. Gehlmann, M. Plötzing, R. Pentcheva, L. Plucinski, C. M. Schneider, M. Müller, *Phys. Rev. B* **2022**, *106*, 054424.

[22] T. S. Santos, J. S. Moodera, K. V. Raman, E. Negusse, J. Holroyd, J. Dvorak, M. Liberati, Y. U. Idzerda, E. Arenholz, *Phys. Rev. Lett.* **2008**, *101*, 147201.

[23] C. Caspers, M. Müller, A. X. Gray, A. M. Kaiser, A. Gloskovskii, C. S. Fadley, W. Drube, C. M. Schneider, *Phys. Rev. B* **2011**, *84*, 205217.

[24] S. G. Altendorf, A. Efimenko, V. Oliana, H. Kierspel, A. D. Rata, L. H. Tjeng, *Phys. Rev. B* **2011**, *84*, 155442.

[25] H. X. Yang, A. Hallal, D. Terrade, X. Waintal, S. Roche, M. Chshiev, *Phys. Rev. Lett.* **2013**, *110*, 046603.

[26] A. Hallal, F. Ibrahim, H. Yang, S. Roche, M. Chshiev, *2D Mater.* **2017**, *4*, 025074.

[27] D. V. Averyanov, I. S. Sokolov, A. M. Tokmachev, O. E. Parfenov, I. A. Karateev, A. N. Taldenkov, V. G. Storchak, *ACS Appl. Mater. Interfaces* **2018**, *10*, 20767.

[28] S. Pandey, S. Hettler, R. Arenal, C. Bouillet, A. R. Moghe, S. Berciaud, J. Robert, J.-F. Dayen, D. Halley, *Phys. Rev. B* **2023**, *108*, 144423.

[29] R. Schiller, W. Müller, W. Nolting, *Phys. Rev. B* **2001**, *64*, 134409.

[30] P. Rosenberger, M. Kundu, A. Gloskovskii, C. Schlueter, U. Nowak, M. Müller, *Sci Rep* **2024**, *14*, 21586.

[31] M. Parschau, K. Christmann, *Surface Science* **1999**, *423*, 303.

[32] M. Jugovac, I. Cojocariu, J. Sánchez-Barriga, P. Gargiani, M. Valvidares, V. Feyer, S. Blügel, G. Bihlmayer, P. Perna, *Advanced Materials* **2023**, *35*, 2301441.

[33] M. DeJarld, P. M. Campbell, A. L. Friedman, M. Currie, R. L. Myers-Ward, A. K. Boyd, S. G. Rosenberg, S. P. Pavunny, K. M. Daniels, D. K. Gaskill, *Sci Rep* **2018**, *8*, 16487.

[34] H. Hibino, H. Kageshima, F. Maeda, M. Nagase, Y. Kobayashi, H. Yamaguchi, *Physical Review B* **2008**, *77*, 075413.

[35] T. J. Regan, H. Ohldag, C. Stamm, F. Nolting, J. Lüning, J. Stöhr, R. L. White, *Phys. Rev. B* **2001**, *64*, 214422.

[36] E. Negusse, J. Dvorak, J. S. Holroyd, M. Liberati, T. S. Santos, J. S. Moodera, E. Arenholz, Y. U. Idzerda, *Journal of Applied Physics* **2009**, *105*, 07C930.

[37] F. Genuzio, T. O. Mentes, A. Locatelli, *IEEE Trans. Magn.* **2019**, *55*, 1.

[38] C. A. Brondin, S. Ghosh, S. Debnath, F. Genuzio, P. Genoni, M. Jugovac, S. Bonetti, N. Binggeli, N. Stojić, A. Locatelli, T. O. Menteş, *Adv Elect Materials* **2024**, *10*, 2300579.

[39] Kie Ahn, G. Almasi, *IEEE Trans. Magn.* **1969**, *5*, 944.

[40] J. J. Yeh, I. Lindau, *Atomic Data and Nuclear Data Tables* **1985**, *32*, 1.

[41] D. A. Siegel, C.-H. Park, C. Hwang, J. Deslippe, A. V. Fedorov, S. G. Louie, A. Lanzara, *Proc. Natl. Acad. Sci. U.S.A.* **2011**, *108*, 11365.

[42] J. L. McChesney, A. Bostwick, T. Ohta, T. Seyller, K. Horn, J. González, E. Rotenberg, *Phys. Rev. Lett.* **2010**, *104*, 136803.

[43] P. Rosenzweig, H. Karakachian, D. Marchenko, K. Küster, U. Starke, *Phys. Rev. Lett.* **2020**, *125*, 176403.

[44] N. J. C. Ingle, I. S. Elfimov, *Phys. Rev. B* **2008**, *77*, 121202.

[45] D. Van Der Marel, G. A. Sawatzky, *Phys. Rev. B* **1988**, *37*, 10674.

[46] A. Locatelli, L. Aballe, T. O. Mentes, M. Kiskinova, E. Bauer, *Surface and Interface Analysis* **2006**, *38*, 1554.

[47] A. Locatelli, T. O. Menteş, C. A. Brondin, I. Cojocariu, M. Jugovac, *Advances in Physics: X* **2025**, *10*, 2549757.

[48] T. O. Menteş, G. Zamborlini, A. Sala, A. Locatelli, *Beilstein Journal of Nanotechnology* **2014**, *5*, 1873.

[49] J. P. Perdew, K. Burke, M. Ernzerhof, *Phys. Rev. Lett.* **1996**, *77*, 3865.

[50] S. Grimme, J. Antony, S. Ehrlich, H. Krieg, *The Journal of Chemical Physics* **2010**, *132*, 154104.

[51] H. Krakauer, M. Posternak, A. J. Freeman, *Phys. Rev. B* **1979**, *19*, 1706.

[52] D. Wortmann, G. Michalicek, N. Baadji, W. Beida, M. Betzinger, G. Bihlmayer, T. Bornhake, J. Bröder, T. Burnus, J. Enkovaara, F. Freimuth, C. Friedrich, C.-R. Gerhorst, S. Granberg Cauchi, U. Grytsiuk, A. Hanke, J.-P. Hanke, M. Heide, S. Heinze, R. Hilgers, H. Janssen, D. A. Klüppelberg, R. Kovacik, P. Kurz, M. Lezaic, G. K. H. Madsen, Y. Mokrousov, A. Neukirchen, M. Redies, S. Rost, M. Schlipf, A. Schindlmayr, M. Winkelmann, S. Blügel, **2024**, DOI 10.5281/ZENODO.7576163.